\documentclass[twocolumn,showpacs,preprintnumbers,amsmath,amssymb,superscriptaddress,floatfix]{revtex4}

\usepackage{graphicx}
\usepackage{dcolumn}
\usepackage{bm}

\begin{document}


\title{Quantum melting of the hole crystal in the spin ladder of Sr$_{14-x}$Ca$_x$Cu$_{24}$O$_{41}$}%

\author{A. Rusydi}
  \affiliation{National Synchrotron Light Source, Brookhaven National Laboratory, Upton, NY, 11973-5000, USA}%
  \affiliation{Materials Science Centre, University of Groningen, 9747 AG Groningen, The Netherlands}%
  \altaffiliation{Current address: Institute for Applied Physics, University of Hamburg, D-20335 Hamburg, Germany}
\author{P. Abbamonte}%
  \affiliation{National Synchrotron Light Source, Brookhaven National Laboratory, Upton, NY, 11973-5000, USA}%
  \affiliation{Physics Department and Frederick Seitz Materials Research Laboratory, University of Illinois, Urbana, IL, 61801}
\author{H. Eisaki}%
  \affiliation{Nanoelectronics Research Institute, AIST, 1-1-1 Central 2, Umezono, Tsukuba, Ibaraki, 305-8568, Japan}
\author{Y. Fujimaki}%
  \affiliation{Department of Superconductivity, University of Tokyo, Bunkyo-ku, Tokyo 113, Japan}
\author{G. Blumberg}%
  \affiliation{Bell Laboratories, Lucent Technologies, Murray Hill, NJ, 07974, USA}
\author{S. Uchida}%
  \affiliation{Department of Superconductivity, University of Tokyo, Bunkyo-ku, Tokyo 113, Japan}
\author{G. A. Sawatzky}%
  \affiliation{Department of Physics and Astronomy, University of British Columbia, Vancouver, B.C., V6T-1Z1, Canada}

\date{\today}

\begin{abstract}
The ``spin ladder" is a reduced-dimensional analogue of the high temperature superconductors that was predicted
to exhibit both superconductivity and an electronic charge density wave or ``hole crystal" (HC).
Both phenomena have been observed in the doped spin ladder system Sr$_{14-x}$Ca$_x$Cu$_{24}$O$_{41}$ (SCCO), which at $x=0$
exhibits a HC which is commensurate at all temperatures.  To investigate the effects of discommensuration
we used resonant soft x-ray scattering (RSXS) to study SCCO as a function of doped hole density, $\delta$. 
The HC forms only with the commensurate wave vectors $L_L = 1/5$ and $L_L = 1/3$ and exhibits a simple temperature scaling 
$\tau_{1/3} / \tau_{1/5} = 5/3$. For irrational wave vectors the HC ``melts", perhaps through the motion of topological 
defects carrying fractional charge.

\end{abstract}


\maketitle

The two-leg ``spin ladder" was introduced\cite{dagotto1} as a computationally more tractable version of the 
{\it t-J} model believed to be relevant to high temperature superconductivity.  
Depending upon the parameters chosen, a
doped ladder can exhibit either
exhange-driven superconductivity\cite{dagotto1,siegrist} or an insulating ``hole crystal" (HC) ground state in which 
the carriers crystallize into a static, Wigner lattice\cite{dagotto1,white,carr}.  
The competition between these two phases is similar to that believed to 
occur between ordered stripes and superconductivity in two-dimensions\cite{jtranAnomaly}.  For this reason, the
spin ladder is an important reference system in the overall understanding of copper-oxides.

The only known doped spin ladder material is Sr$_{14-x}$Ca$_x$Cu$_{24}$O$_{41}$ (SCCO), which for $x>10$
was shown remarkably to exhibit superconductivity under pressure\cite{uehara,kojima}.  There is now evidence from 
transport\cite{tomic1,girshScience}, Raman scattering\cite{girshScience}, and resonant x-ray scattering\cite{peteNature} 
that under ambient conditions the holes lie in the competing crystallized state, i.e. form a ``hole crystal" (HC).
At $x=0$ this HC was shown in Ref. \cite{peteNature} to be commensurate at all temperatures.  
This contrasts with the behavior of most Peierls charge density waves (CDWs) in which the wavelength
typically shifts and becomes commensurate only at low temperatures\cite{gruner}.  This raises the question of
what role commensuration plays, in general, in the physics of a Wigner crystal (WC).

Commensuration effects on a WC were originally considered by Hubbard\cite{hubbard}, 
in the context of conducting polymers, who showed that a WC is most stable when the carrier density is a 
rational fraction and its wavelength commensurate with the lattice.  
Excess carriers added were postulated to form topological defects or ``solitons" which carry fractional 
charge\cite{hubbard,su,rice}.  
At a sufficient degree of incommensurability, i.e. soliton density, 
the system becomes unstable to the formation of additional soliton/anti-soliton pairs and the WC melts.  This
``incommensurate melting" is a purely quantum mechanical effect and is fundamentally different from 
discommensuration effects exhibited by a Peierls-type CDW\cite{gruner}.  
The CDWs in polymers, such as polyacetylene 
or TCNQ salts, were later shown to be of the Peierls type \cite{itsPeierls}.
To investigate the effects of commensuration on a real WC, we have used 
resonant soft x-ray scattering (RSXS) \cite{peteScience,duur,wilkins,sanjeet,thomas,schusslerNickelate}
to study SCCO as a function of doped hole density, $\delta$.

SCCO is an adaptive misfit material consisting of chain and ladder
subsystems with incompatible periods $\sqrt{2} \, c_c \approx c_L = 3.90 \AA$.  The misfit results
in a buckling along the $c$ axis, sometimes interpreted as a unit cell, 
with period $c \approx 10 c_c \approx 7 c_L$, though the period is frequently non-integer\cite{hiroi}.  
A CDW was reported in the chain layer\cite{matsuda1,matsuda2,cox,fukuda}, 
though further examination has raised suspicion that this effect is related to the buckling\cite{vansmaalen,etrillard}.

SCCO is self-doped with 6 holes per formula unit, most of which reside in the chain because of its lower
Pauling electronegativity.  At $x=0$ approximately five of the six holes reside in the chain\cite{nucker,osafune,magishi}.  
Substitution of Ca for Sr transfers holes from the chain to the ladder\cite{nucker, osafune, magishi}, 
though the magnitude of this transfer is unclear.  X-ray absorption measurements\cite{nucker}
suggest that $\delta$ for the ladder ranges from 0.057 for $x$ = 0 to 0.079 for $x$ = 12.  Optical conductivity studies
suggest a range from 0.07 to 0.2\cite{osafune} while $^{63}$Cu NMR studies suggest a range 0.07 to 0.25\cite{magishi}.  
The lack of consensus poses a problem for an $x$-dependent study of
SCCO.  However it is agreed that the transfer is linear in $x$ and we will show that much can be learned using only this.

Single crystals of Sr$_{14-x}$Ca$_x$Cu$_{24}$O$_{41}$ (SCCO) with {\it x} = 0, 1, 2, 3, 4, 5, 10, 11, and 12 were
grown by high-pressure floating zone techniques.  
Crystals were cut with a wire saw
and dry polished with diamond film down to a grit of 0.05 $\mu$m.  
Resonant soft x-ray scattering studies were
carried out on beam line X1B at the National Synchrotron Light Source with a ten-axis, ultrahigh vacuum-compatible
diffractometer.  We denote periodicity with Miller indices of the ladder, i.e.
$(H,K,L_L)$ denotes a momentum transfer ${\bf Q} = (\frac{2\pi}{a} H, \frac{2\pi}{b} K, \frac{2\pi}{c_L} L_L)$
where $a = 11.47 \AA$, $b = 13.35 \AA$, $c_L=3.90\AA$.  

In Ref. \cite{peteNature} we reported that SCCO, for $x=0$,  contains a HC with commensurate 
wave vector $L_L = 1/5$.  Here we report the observation of another HC 
reflection for $x$ = 10, 11, 12 with commensurate wave vector $L_L = 1/3$ (Fig. 1).  
The higher wave vector (shorter period) observed makes sense given the higher values of $x$ and $\delta$.  
Like the phenomenon at $x=0$ this reflection breaks the superspacegroup symmetry of the 
composite unit cell \cite{etrillard} so cannot be a reflection from the chain-ladder buckling.  
In contradiction with previous transport and Raman 
studies\cite{adrianDoping,vuleticDoping} no hole crystal was observed at any of the intermediate dopings $1 \leq x \leq 5$.
The intensity of the HC as a function of Ca content for the full range
is summarized in Fig. 1.  Reminiscent of previous proposals\cite{hubbard,su,rice,white}, the HC forms only with commensurate wave vector.
The doping is not precisely known but
presumably these coincide with rational values of the hole density, $\delta$.  

Assuming $\delta$, $x$ and the wave vector $L_L$ are all linearly related Fig. 1 suggests the relationship

\begin{equation}
L_L = 1/5 + 2 x / 165.
\end{equation}

\noindent
From this one might expect a HC with wave vector $L_L = 1/4$ to form at $x = 4.125$, however
none is observed.  It appears that the HC is stable for odd, though not even, multiples of the ladder period.  

The stabilizing effect of commensurability is best seen in Fig. 2 which shows reciprocal space maps in the $(H,0,L_L)$ plane
around the position $(0,0,1/3)$ for $x = 10, 11, 12$ ($T=20K$).  The maximum HC intensity 
occurs at $x=11$ where the wave vector is closest to 1/3.  As $\delta$ is varied
the HC shifts slightly from the commensurate position and ``melts", reminiscent of the soliton mechanism postulated for
organic chains\cite{hubbard,rice,su}.  It is not clear how to tell if solitons are present,
but the width of the $L_L=1/3$ peak in Fig. 1, 
$\Delta L_L = 0.012$, would suggest a critical soliton spacing of $\sim 82 c_L$.  

The HC correlation length may be read 
off the maps in Fig. 2 to be isotropic with the value $\xi_{1/3} = (1600 \pm 90) \AA 
\sim 410 c_L$.
This should be compared with the $\xi_{1/5} = 255\AA \sim 65 c_L$ reported at $x=0$\cite{peteNature} 
($\xi$ is temperature-independent in all samples).  That $\xi$ is larger at larger $L_L$ shows 
that the interactions which drive
the HC get stronger as the distance between holes decreases.

The reflection with wave vector $L_L=1/5$ reported at $x=0$ \cite{peteNature} was visible 
only with the incident x-ray energy tuned to
the mobile carrier prepeak (MCP) below the O$K$ edge and the ligand hole sideband of the Cu$L_{3/2}$
edge.  This unusual spectroscopic signature was the reason for identifying this phenomenon as a HC.
Slightly different effects are seen for $L_L=1/3$ in $x=10,11,12$.

In Figure 3 we display a ``resonance profile" (RP) - the intensity of the HC scattering vs. incident energy 
compared to x-ray absorption spectra (XAS).  
XAS was taken {\it in situ} in fluorescence yield mode and is consistent with 
previous studies\cite{nucker}.  RPs are shown at both edges for $x=0,10,11,12$.  
Like the case of $L_L=1/5$, the $L_L=1/3$ HC was visible only when the 
x-ray energy was tuned to oxygen and copper thresholds\cite{peteNature}
suggesting a primarily, though not necessarily exclusively, electronic origin\cite{finePrint}.  
The O$K$ RP for the $x=10$ and $x=12$ samples is identical to that at 
$x=0$; the HC scattering resonates at the ladder shoulder of the MCP.  
By contrast, the MCP resonance at $x=11$, where the HC scattering is 
strongest, is slightly red-shifted compared to the others and exhibits a second resonance at the O$K$
edge jump at 532.7 eV.  This suggests that the hole modulation at $x=11$ is large enough to modulate the O$2p$
continuum and cause a slight core level shift.  We did not quantify the hole amplitude but most of the ladder
holes are probably modulated at this doping.  

New behavior is observed for $L_L=1/3$ at the Cu$L_{3/2}$ edge.  Unlike at $x=0$,
the RP for $x=10,11,12$ consists of two peaks - one at 930 eV, likely due to the distortions in the Cu sublattice, and a 
second at 933 eV due to the hole modulation itself.  The lineshape arises from coherent interference between the two 
and can be used, in principle, to determine the phase between the hole and structural modulations.  
This will be the subject of a forthcoming article.  
Overall, it is evident that that the HC at $10 \leq x \leq 12$ is mainly electronic
but coupled to the lattice more strongly than at $x=0$. 

The temperature dependence of the HC reveals some character of the interactions which drive it.  Figure 4 shows the integrated
intensity of the HC peak vs. $T$ for $x=0,10,11,12$.  While the $L_L=1/5$ HC vanishes by $T = 300$ K, the
$L_L=1/3$ HC persists to much higher temperature ($T > 300K$ was not attempted for fear of sample deoxygenation).  Like the 
larger coherence length, this indicates that the interactions driving the HC are stronger i.e. increase 
with decreasing spacing between holes.  
If plotted against the reduced quantities $I/I_{20K}$ and $T/\tau_{L_L}$, 
where $\tau_{1/3}=211.1 K$ and $\tau_{1/5}=127.8 K$, the data collapse to a universal curve (Fig. 4, inset).  
The ratio of the 
temperature scales $\tau_{1/3}/\tau_{1/5} = 1.65 = 5/3$, or in other words the energy scale of the HC varies inversely with its
wavelength.  This apparent $1/r$ dependence suggests that 
the HC is driven simply by direct, long-range Coulomb repulsion.  This is not the RVB interaction
usually considered for ladders\cite{white,dagotto1} but is what was originally used by Hubbard\cite{hubbard} and explains why the
coherence length $\xi$ is isotropic despite the anisotropic bonding environment.  

It is important to address the relationship between our measurements and the charge-order reported 
previously\cite{matsuda1,matsuda2,cox,fukuda,vansmaalen,etrillard}.  We have observed the lowest-order such reflection,
which occurs at $L_L=0.45$ or $L_c=0.31$ in chain units and was identified as a chain-ladder buckling reflection by Hiroi\cite{hiroi}.  
This peak has little temperature dependence but has a mixed structure/hole character suggesting the chain holes are 
modulated by the buckling strain wave.  This effect will be described in detail in a forthcoming article.

It is unsettling that there is disagreement between our results 
and Raman measurements\cite{adrianDoping}, which report a HC for all $x$, and 
impedence measurements\cite{vuleticDoping}, which report a HC for $x<9$.  
Since these measurements contradict one another reconciliation with both is impossible.
We note, however, that there is direct correspondence between the HC in our measurements 
and the presence of a metal insulator transition under pressure\cite{kojima}.

Our results clearly demonstrate that commensuration is critically important for the formation of the hole crystal in SCCO.  
The quantum melting observed for incommensurate wave vectors might indicate the presence of solitons with fractional 
charge\cite{hubbard,su,rice}.  We see strong evidence for the role of unscreened Coulomb interactions, which 
should be expected in any material in which the valence electrons are (self-consistently) localized.  
We speculate that such effects may also be responsible for 
the static spin/charge order and suppression of $T_c$ seen other cuprates near $\delta=1/8$\cite{jtranAnomaly,fujitaAnomaly}.  

We acknowledge helpful discussions with Ian Affleck, Wei Ku, Alexei Tsvelik, and Antonio Castro-Neto.  This work was supported by the 
U.S. Department of Energy under contract DE-AC02-98CH10886, the Netherlands Organization for Fundamental Research on Matter (FOM), 
and the 21st Century COE program of the Japan Society for Promotion of Science.

\begin{figure}
\includegraphics{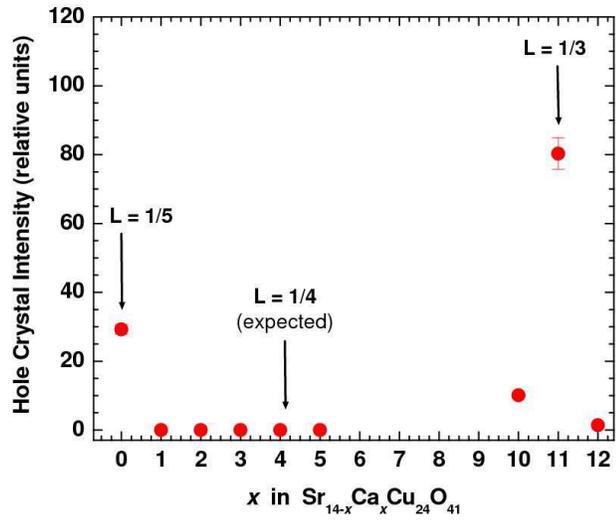}
\caption{
Intensity of hole crystal scattering at the O$K$ resonance maximum for all chemical compositions studied.  
Hole crystallization occurs only with the rational wave vectors $L_L=1/5$ and $L_L=1/3$.  No reflection with 
$L_L=1/4$ is observed.  
}
\end{figure}

\begin{figure}
\includegraphics{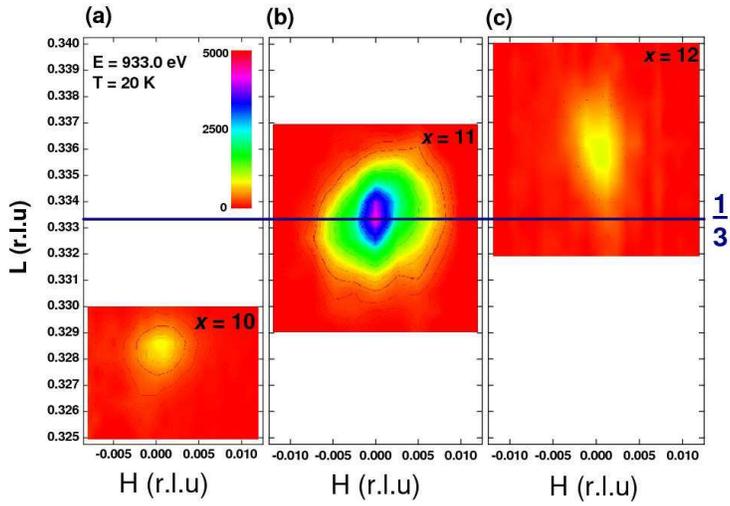}
\caption{
Reciprocal space maps around $(H,0,L_L) = (0,0,1/3)$ at the Cu$L_{3/2}$ resonance (photon energy 933 eV) for 
(a) $x = 10$, (b) $x = 11$ and (c) $x =12$.  The HC is most pronounced when its wave vector has the commensurate 
value $L_L = 1/3$.  As $L_L$ shifts with the addition of carriers the HC ``melts".  
All maps were taken with T=20K.
}
\end{figure}

\begin{figure}
\includegraphics{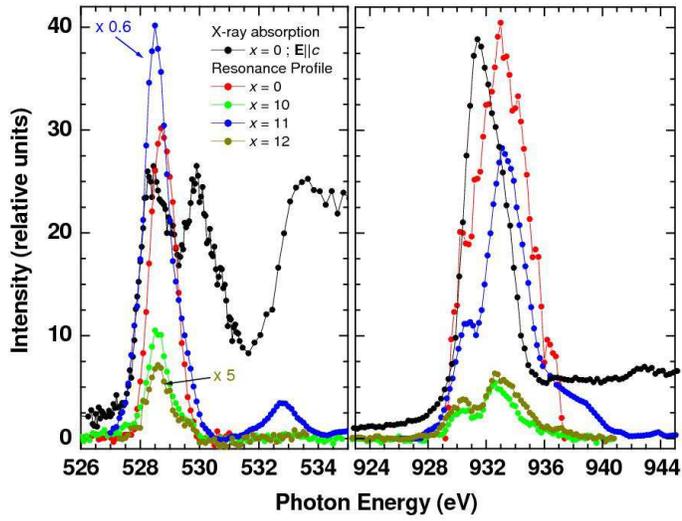}
\caption{
``Resonance profiles", i.e. the integrated intensity of HC scattering as a function of incident x-ray energy, compared to 
x-ray absorption spectra (XAS) of the $x=0$ system (${\bf E}||c$) near (a) the O$K$ and (b) the Cu$L_{3/2}$ edges.  The XAS spectrum 
is shown in black.  Red, green, blue, and brown symbols correspond to $x$ = 0, 10, 11 and 12 respectively.
}
\end{figure}

\begin{figure}
\includegraphics{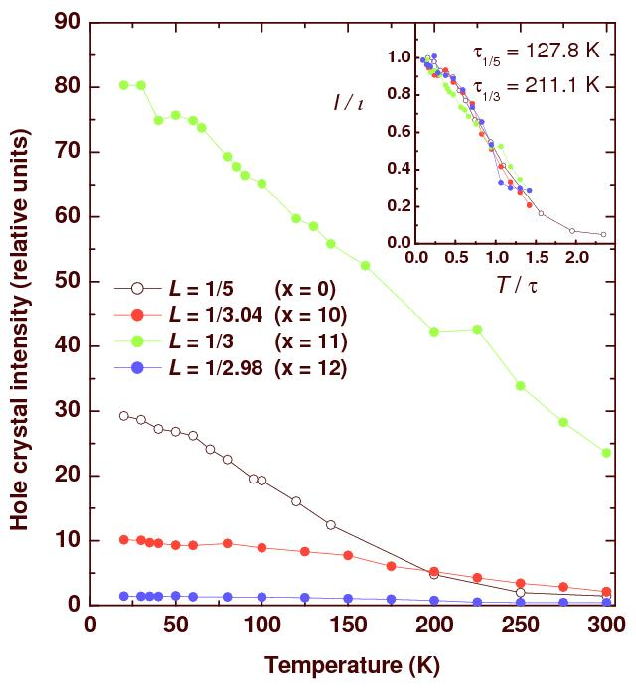}
\caption{
Temperature dependence of HC scattering for the four samples in which it has been observed. 
Open, red, green, and blue circles correspond to samples with x = 0, 10, 11 and 12, respectively.  
The data collapse when plotted against the reduced quantities $I/I_{20K}$ and $T/\tau_{L_L}$, 
where $\tau_{1/3}=211.1 K$ and $\tau_{1/5}=127.8 K$.
}
\end{figure}

\bibliography{comm}

\begin{thebibliography}{35}
\bibitem[{fin()}]{finePrint}
\bibinfo{note}{Nonresonant scattering from structural distortions is innately
  weak in the soft x-ray regime because both the scattering volume and the
  factor $({\bf Q}\cdot\Delta)^2$ are small. Even considering these two
  efffects, however, the off-resonant scattering here is anomalously weak.}
\end{thebibliography}

\end{document}